\documentstyle[11pt,epsfig,amssymb]{article}
\title{\bf Brans-Dicke DGP Brane Cosmology}
\author{K. Atazadeh\thanks{email: k-atazadeh@sbu.ac.ir}\,\ and H. R.
Sepangi\thanks{email: hr-sepangi@sbu.ac.ir}
\\ {\small Department of Physics, Shahid Beheshti University, Evin,
Tehran 19839, Iran}}
\begin{document}
\maketitle
\begin{abstract}
We consider a five dimensional DGP-brane scenario endowed with a
non-minimally coupled scalar field within the context of Brans-Dicke
theory. This theory predicts that the mass appearing in the
gravitational potential is modified by the addition of the mass of
the effective intrinsic curvature on the brane. We also derive the
effective four dimensional field equations on a $3+1$ dimensional
brane where the fifth dimension is assumed to have an orbifold
symmetry. Finally, we discuss the cosmological implications of this
setup, predicting an accelerated expanding universe with a value of
the Brans-Dicke parameter $\omega$ consistent with values resulting
from the solar system observations.
 \end{abstract}

\section{Introduction}

Over the past decade the possibility of the observable Universe
being a brane-world \cite{Rubakov} embedded in a higher dimensional
space-time has generated a great amount of interest. This was
motivated by the fact that there is a strongly coupled sector of
$E_8 \times E_8$ heterotic string theory which can be described by a
field theory living in 11-dimensional space-time \cite{horava}. The
11-dimensional world consists of two 10-dimensional hypersurfaces
embedded on the fixed points of an orbifold and the matter fields
are assumed to be confined and live on these hypersurfaces which are
known to be 9-branes. After compactification of the 11-dimensional
theory on a Calabi-Yau 3-fold, we obtain an effective 5-dimensional
theory \cite{lukas} which has the structure of two 3-branes located
on the orbifold boundaries. This scenario has motivated intense
efforts to understand the case where the bulk is a 5-dimensional
anti de-Sitter space. In this setup, gravitons are allowed to
penetrate into the bulk but are localized on and around the brane
\cite{lisa}. It was then shown that in a background of a
non-factorizable geometry an exponential warp factor emerges which
multiplies the Poincar\'e invariant 3+1 dimensions in the metric.
The model consists of two $3$-branes situated along the 5th
dimension, compactified on a $S^{1}/Z_2$ orbifold symmetry where the
two branes must have opposite tensions. The evolution equation
followed from such a brane scenario differs from  that of the
standard four dimensional evolution equation when no branes are
present~\cite{bin}. The existence of branes and the requirement that
matter fields should be localized on the brane lead to a
non-conventional cosmology, necessitating a more concerted study. A
large number of studies have been devoted to the effective gravity
induced on the brane~\cite{brane} and, in particular, a great amount
of interest was generated on inflationary cosmology~\cite{numero}.
More recently, post-inflationary brane cosmology has been also
considered in ~\cite{anu}. Not surprisingly, the problem of the
cosmological constant has become a focal point in the brane-world
studies where, for example, in \cite{nim,kachru,bin1} a five
dimensional action with a scalar field is non-minimally coupled to
five dimensional gravity and to the four dimensional brane tension.
There has also been some discussion on the localization of gravity
\cite{gomez}. A feature common to these type of models is that they
predict deviations from the usual $4D$ gravity at short distances.

A somewhat different approach within the brane-world framework is
the model proposed by Dvali, Gabadadze and Porrati (DGP)
\cite{dvali,dvali2}. It predicts deviations from the standard $4D$
gravity over large distances. The transition between four and
higher-dimensional gravitational potentials in the DGP model arises
because of the presence of both the brane and bulk Einstein terms in
the action. The Friedmann-like equations governing the cosmological
evolution of a brane possessing an intrinsic curvature term in its
action have already been derived and discussed for an
AdS-Schwarzschild bulk space-time \cite{holdom}. Cosmological
consideration of the DGP model was first discussed in
\cite{deffayet} where it was shown that in a Minkowski bulk space
time we can obtain self-accelerating solutions. In the original DGP
model it is known that $4D$ general relativity is not recovered at
linearized level. However, some authors have shown that at short
distances we can recover the $4D$ general relativity in a
spherically symmetric configuration, see for example \cite{tanaka}.
An important observation was made in \cite{dick,dick2} where it was
shown that the DGP model allows for an embedding of the standard
Friedmann cosmology whereby the cosmological evolution of the
background metric on the brane can entirely be described by the
standard Friedmann equation plus energy conservation on the brane.
This was later extended to arbitrary number of transverse dimensions
in \cite{Cordero}. For a comprehensive review of the phenomenology
of DGP cosmology, the reader is referred to \cite{Lue}.

It is worth mentioning that an interesting feature of the original
DGP model is the existence of ghost-like excitations \cite{ghost}.
We do not endeavor to discuss such ghost excitation in the present
study since our aim is the study of the cosmological implications of
the model presented in this work. For a comprehensive review of the
existence of $4D$ ghosts on the self-accelerating branch of
solutions in DGP models see \cite{gregory}.

In this paper we consider the dynamics of a scalar field existing in
the bulk and study the cosmological implications that such a
scenario would entail in a DGP brane-world. We do this by studying
the effective field equations on the $3+1$ dimensional DGP brane
which is assumed to be rigidly located on the orbifold symmetry
along the 5th dimension. In section 3 we consider the weak field
approximation and show that this theory predicts a modification to
the mass appearing in the gravitational potential in the form of an
additional mass term related to the effective intrinsic curvature
and that such a mass can be interpreted as dark matter. Finally, in
section 4, we discuss the ensuing cosmology, showing that the model
predicts an accelerated expanding universe without introducing dark
energy and is consistent with the present observational bounds on
the value of the Brans-Dicke parameter $\omega$. Conclusions are
drawn in the last section.


\section{Brans-Dicke Brane in DGP Scenario}

Let us start by considering a bulk scalar field non-minimally
coupled to gravity in a 5-dimensional DGP brane world. Such a
scalar field in the five dimensional theory can be viewed as a
dilaton which is purely an outcome of dimensional reduction from
some higher dimensional theory to a 5-dimensional space-time
\cite{mendes}. The action for our model can be written as
\begin{equation}
\label{action} S_{5}=\frac{1}{2 \kappa_{(5)}^2}\int d^5x
\sqrt{-g}\left(\phi{\cal R} - \frac{\omega}{\phi}\partial_{A}\phi
\partial ^{A} \phi \right) +\frac{1}{2\mu^{2}}\int
d^{4}x\sqrt{-q}R\Big|_{\rm brane}+ \int d^{5}x \sqrt{-g} {\cal
L}_{m}\,,
\end{equation}
where ${\cal R}$ is the Ricci scalar associated with the
5-dimensional space-time metric $ g_{_{AB}}$, $\phi$ is a scalar
field which we shall call the Brans-Dicke (BD) field, $\omega$ is a
dimensionless coupling constant which determines the coupling
between gravity and the BD scalar field. Similarly, the second term
is the Einstein–-Hilbert action for the induced metric $q_{\mu\nu}$
on the brane where $R$ is the relevant scalar curvature,
$\mu^{2}=8\pi G_{(4)}$ and  ${\cal L}_{\rm m}$ represents the
Lagrangian for the matter fields. Latin indices denote 5-dimensional
components ($A,B=0,\ldots,5$) and for convenience we choose
$\kappa_{(5)}^2=8\pi G_{(5)}=1 $. The induced metric $q_{\mu\nu}$ is
defined as usual from the bulk metric $g_{_{AB}}$ by
\begin{equation}\label{eq1}
q_{\mu\nu} = \delta^{A}_{\mu} \delta^{B}_{\nu }g_{_{AB}}.
\end{equation}
The variational derivative of the action equation (\ref{action})
with respect to $g_{_{AB}}$ yields the field equations
\begin{eqnarray}\label{eq2}
G_{AB}\equiv{\cal R}_{AB}- \frac{1}{2}  g_{_{AB}} {\cal R} & = &
\frac{1}{\phi}\left[ T_{AB}^{\phi}+T^{
curv}_{AB}+T_{AB}\right],
\end{eqnarray}
where
\begin{eqnarray}\label{eq3}
T_{AB}^{\phi}=\frac{\omega}{\phi} \left[\phi_{;A}\phi_{; B}
-\frac{1}{2}  g_{_{AB}}\phi_{; C}\phi^{;
C}\right]+\left[\phi_{_{;AB}}- g_{_{AB}} {\phi^{;C}}_{;C} \right]
\,,
\end{eqnarray}
and
\begin{eqnarray}\label{curv}
T^{curv}_{AB}=-\frac{1}{\mu^{2}}g^{\mu}_{_{A}}g^{\nu}_{_{B}}\left[R_{\mu\nu}-\frac{1}{2}q_{\mu\nu}R\right].
\end{eqnarray}
Note that $T^{ curv}_{AB}$ is the contribution coming from the
scalar curvature of the brane. The equation of motion for the scalar
field $\phi$ is given by
\begin{eqnarray}\label{eq4}
\square \phi  & = & \frac{(T+T^{ curv})}{3\omega +4}\,,
\end{eqnarray}
where ${ T= T^{C}}_{C}$ is the trace of the energy-momentum tensor
of the matter content of the $5$-dimensional space-time. Notice the
factor $3\omega+4$ in the denominator on the right hand side of the
BD field equation instead of the familiar $2\omega+3$ in the
$4$-dimensional case~\cite{barrow}. This is determined by requiring
the validity of the equivalence principle in our setup, see
\cite{wein} for a discussion of this topic in the context of
$4$-dimensional Brans-Dicke theory.

\section{Brans-Dicke DGP brane model in the weak field approximation regime}
Although equations (\ref{eq2}) and (\ref{eq4}) represent the more
usual or standard form of Brans-Dicke equations in the DGP model but
to consider the weak field approximation of the model we shall work
in the Einstein frame \cite{dicke,romero} given by
\begin{eqnarray}
\label{2.4} \tilde G_{AB} = {\cal G} \left[\tilde T_{AB}+
\frac{3\omega+4}{2{\cal G}\phi^{2}} \left(\phi_{; A}\phi_{; B}
-\frac{1}{2} \tilde{g}_{_{AB}}\phi_{; \rm C}\phi^{;C}\right)+\tilde
T^{curv}_{AB}\right],
\end{eqnarray}
\begin{eqnarray}
\label{2.5} \tilde \square\ln({\cal G} \phi) ={{\cal G}\over
3\omega +4}[\tilde T+\tilde T^{\it curv}],
\end{eqnarray}
which is obtained from (\ref{eq2}) and (\ref{eq4}) by making the
transformation
\begin{eqnarray}
\label{2.6} \tilde g_{_{AB}}={\cal G}\phi g_{_{AB}},
\end{eqnarray}
\begin{eqnarray}
\label{2.7} \tilde T_{AB}={\cal G}^{-1}\phi^{-1}
T_{AB}~~~~~~~~~\mbox{and}~~~~~~~~~~\tilde T^{ curv}_{AB}={\cal
G}^{-1}\phi^{-1} T^{ curv}_{AB},
\end{eqnarray}
where ${\cal G}$ is an arbitrary constant and the tilde on $\tilde
G_{AB}$, $\tilde\square$, $\tilde T^{\it curv}$ and $\tilde T$ means
that these quantities are calculated using the conformal metric
$\tilde g_{_{AB}}$.

In the weak field approximation of Brans-Dicke theory, in addition
to
\begin{eqnarray}\label{per}
 g_{_{AB}} =\eta_{_{AB}} + h_{_{AB}},
\end{eqnarray}
we must also assume that
\begin{eqnarray}
\label{2.8} \phi = \phi^{(0)} + \phi^{(1)},
\end{eqnarray}
where $\phi^{(1)} = \phi^{(1)}(x^{\mu},y)$ is a first-order term in
the energy density and $\left|{\phi^{(1)} \over \phi^{(0)}} \right|
\ll 1$. Taking into account (\ref{2.8}) and setting ${\cal G}=\frac
{1}{\phi^{(0)}}$ the transformation equations (\ref{2.6}) and
(\ref{2.7}) become
\begin{eqnarray}
\label{2.9} \tilde g_{_{AB}}=\eta_{_{AB}}+\tilde h_{_{AB}},
\end{eqnarray}

\begin{eqnarray}\label{2.10}
\tilde T_{AB}=(1-\phi^{(1)} {\cal G})T_{AB}=T_{AB}
\end{eqnarray} and
\begin{eqnarray}\label{2.111}
\tilde T^{\it curv}_{AB}=(1-\phi^{(1)}{\cal G})T^{\it curv}_{AB}=T^{\it curv}_{AB},
\end{eqnarray}
where
\begin{eqnarray}
\label{2.11} \tilde h_{_{AB}}= h_{_{AB}}+\phi^{(1)}{\cal
G}\eta_{_{AB}},
\end{eqnarray}
and only the first-order terms in the mass and curvature densities
have been retained. Note that $T^{\it curv}_{AB}$ lives on the
brane. Now, substituting (\ref{2.8}) in the field equations
(\ref{2.4}) and taking equations (\ref{2.9}) and  (\ref{2.10})
into account, we get
\begin{eqnarray}
\label{2.12} \tilde G_{AB}={\cal G} [T_{AB}+T^{ curv}_{AB}].
\end{eqnarray}
On the other hand, the scalar field equation (\ref{eq4}) becomes

\begin{eqnarray}
\label{2.13} \square \phi^{(1)} = {[T+T^{\it curv}]\over 3\omega +4}.
\end{eqnarray}
It turns out then that equations (\ref{2.12}) are formally identical
to the field equations of General Relativity with ${\cal G}$
replacing the Newtonian gravitational constant $G_{(5)}$, dropping
the coefficient $8\pi$. Therefore, if $\tilde
g_{_{AB}}(G_{(5)},x^{\mu},y)$ is a known solution of the Einstein
equations in the weak field approximation for a given $T_{AB}$, then
the Brans-Dicke solution corresponding to the same $T_{AB}$ will be
given in the weak field approximation just by taking the inverse of
equation (\ref{2.6}), that is

\begin{eqnarray}
\label{2.14} g_{_{AB}}(x^{\mu},y)={\cal G}^{-1}\phi^{-1}\tilde
g_{_{AB}}({\cal G},x^{\mu},y)=\left[1-\phi^{(1)}(x^{\mu},y){\cal
G}\right]\tilde g_{_{AB}}({\cal G},x^{\mu},y),
\end{eqnarray}
or, equivalently,
\begin{eqnarray}
\label{2.15} h_{_{AB}}(x^{\mu},y)=\tilde h_{_{AB}}({\cal G}
,x^{\mu},y)-\phi^{(1)}(x^{\mu},y){\cal G}\eta_{_{AB}}.
\end{eqnarray}
We therefore conclude that the general problem of finding solutions
of Brans-Dicke DGP equations of gravity in the weak field
approximation may be reduced to solving Einstein field equations for
the same matter distribution. It should be noted that the Einstein
tensor $\tilde G_{AB}$ which appears on the left hand side of
(\ref{2.12}) must be calculated in the weak field approximation,
i.e., taking $\tilde g_{_{AB}}$ as given by (\ref{2.9}).

Now, it is well known from equation (\ref{2.12}) that the
gravitational potential in the weak field approximation of DGP
expressed in Gaussian normal coordinates in the Einstein frame is
given by \cite{dick2}
\begin{equation}\label{poten}
\tilde U(\vec{r})=-\frac{\mu^{2}M}{6\pi r}
\left[\cos(2\mu^{2}r)-\frac{2}{\pi}\cos(2\mu^{2} r) {\rm
si}(2\mu^{2}r)+\frac{2}{\pi}\sin(2\mu^{2}r) {\rm
ci}(2\mu^{2}r)\right],
\end{equation}
where the sine and cosine integrals are defined by the following
relations.
\begin{eqnarray}\label{si}
{\rm
si}(x)=\int_{0}^{x}d\xi\frac{\sin\xi}{\xi}~~~~~~\mbox{and}~~~~~~~~~{\rm
ci}(x)=-\int_{x}^{\infty}d\xi\frac{\cos\xi}{\xi}.
\end{eqnarray}
Note that this equation is  the gravitational potential for the
mass density $\rho(\vec{r})=M\delta(\vec{r})$. We must now solve
equation (\ref{2.13}), leading to $\phi^{(1)}$ given by
\begin{equation}\label{phi}
\phi^{(1)}=-\frac{M+M^{\it curv}}{4\pi(3\omega+4)r}.
\end{equation}
Using equations (\ref{2.11}), (\ref{poten}) and (\ref{phi})
gravitational potential in Jordan frame  given by
\begin{equation}\label{poten2}
U(\vec r)=\tilde U(\vec r)-\frac{{\cal G}(M+M^{\it curv})}{4\pi(3\omega+4)r},
\end{equation}
where it seems as if $M^{\it curv}$ plays the role of dark matter
which has a contribution to the gravitational potential. It
therefore seems plausible that $\rho^{\it curv}(\vec{r})=M^{\it
curv}\delta(\vec{r})$ could be considered as a candidate for dark
matter. From equation (\ref{poten2}) we see that the Brans-Dicke DGP
model predicts a transition scale relating the $4$ and
$5$-dimensional behavior of the gravitational potential in the
Jordan frame
\begin{eqnarray}\label{potenlimit}
\left\{
\begin{array}{lll}
r\ll\ell_{_{DGP}}:~U(\vec{r})=-\frac{4\mu^2M(3\omega+4)+6{\cal
G}(M+M^{curv})}{24\pi(3\omega+4)r}
-\frac{\mu^{2}M}{6\pi r}\left[
\left(\gamma-\frac{2}{\pi}\right)\frac{r}{\ell_{_{DGP}}}+
\frac{r}{\ell_{_{DGP}}}\ln\left(\frac{r}{\ell_{_{DGP}}}\right)+{\cal
O}\left(\frac{r}{\ell_{_{DGP}}}\right)^{2}\right],\\\nonumber\\\nonumber
r\gg\ell_{_{DGP}}:~U(\vec{r})=-\frac{{\cal G}(M+M^{\it
curv})}{4\pi(3\omega+4)r}-\frac{\mu^{2}M}{6\pi^{2}r^{2}}-
\frac{\mu^{2}M}{6\pi^{2}r^{2}}\left[-2\frac{\ell_{_{DGP}}^{2}}{r^{2}}+
{\cal O}\left(\frac{\ell_{_{DGP}}}{r}\right)^{4}\right].
 \end{array}
\right.
\end{eqnarray}
Here $\ell_{_{DGP}}=\frac{\mu^{2}}{2}$ and $\gamma\simeq0.577$ is
Euler's constant .
\section{Cosmological considerations}
Before we discuss the energy-momentum tensor, let us define the five
dimensional metric which has the following form
\begin{eqnarray}
\label{met0} ds^2= g_{_{\rm AB}}dx^{\rm A}dx^{\rm B}=q_{\mu
  \nu}(x^{\mu},y)\, dx^{\mu}dx^{\nu} + b^2(x^{\mu},y) \, dy^2 \, ,
\end{eqnarray}
where $\mu, \nu =0,\ldots,3$ and $y$ is the coordinate associated
with the fifth dimension and we will adopt a brane-based approach
where the brane is the hypersurface defined by $y=0$. We also
assume an orbifold symmetry along the fifth direction $y
\rightarrow -y$. As we shall see in the coming sections, this will
help us to simplify our calculations. Next we define the
energy-momentum tensor
\begin{eqnarray}
\label{em0} {T^{\rm A}}_{\rm B}= {T^{\rm A}}_{\rm B}\arrowvert_{\rm
bulk} + {T^{\rm A}}_{\rm B} \arrowvert_{\rm brane}\,,
\end{eqnarray}
where the subscripts ``brane'' and ``bulk'' refer to the
corresponding energy-momentum tensors. For simplicity we assume
that the bulk is devoid of matter other than the Brans-Dicke
scalar field. The brane matter field is held at $y=0$ with the
following energy momentum tensor
\begin{eqnarray}
\label{em1} {T^{\rm A}}_{\rm B}\arrowvert_{\rm brane}=
\frac{\delta(y)}{b}{\rm diag}(-\rho,p,p,p,0)~~~~~~~~\mbox{and}~~~~~~~~{T^{\rm A}}_{\rm B}\arrowvert_{\rm bulk}={\rm diag}(0,0,0,0,0).
\end{eqnarray}
The above expressions are written assuming that the brane is thin
and that the bulk is empty.

Since we are interested in exploring the spatially flat cosmology
($k=0$), we consider a $5$-dimensional flat metric {\it ansatz} of
the following form
\begin{eqnarray}
\label{met1}
ds^2=-n^2(\tau,y)d\tau^2+a^{2}(\tau,y)\delta_{ij}dx^{i}dx^{j} +
b^2(\tau,y) dy^2\,,
\end{eqnarray}
where $i,j =1,2,3$. With this metric we are now able to write the
equations of motion. The $(0,0)$ component reads
\begin{eqnarray}
\label{eq:00} 3 \left[\frac{\dot a}{a}\left(\frac{\dot
a}{a}+\frac{\dot b}{b}\right)
 - \frac{n^2}{b^2} \left(\frac{a^{\prime
\prime}}{a}+\frac{a^{\prime}}{a}\left(\frac{a^{\prime}}{a}-
\frac{b^{\prime}}{b}\right)\right)\right] &=&
\frac{1}{\phi}\left[T^{\phi}_{\rm 00}+T_{00}+T^{curv}_{\rm
00}\right] \, ,
\end{eqnarray}
where
\begin{eqnarray}\label{Teq:00}
 T_{\rm 00}^{\phi}&=&-\dot{\phi}\left(3\frac{\dot{a}}{a} +
\frac{\dot{b}}{b} -\frac{\omega}{2} \frac{\dot{\phi}}{\phi}\right) +
\left( \frac{n}{b} \right)^2 \left[ \phi^{\prime\prime} +
\phi^{\prime} \left( 3 \frac{a^{\prime}}{a} - \frac{b^{\prime}}{b} +
\frac{\omega}{2} \frac{\phi^{\prime}}{\phi} \right)\right]
\end{eqnarray}
and
\begin{equation}\label{Tcurveq:00}
T^{\it curv}_{\rm
00}=-\frac{3}{\mu^{2}b}\left(\frac{\dot{a}}{a}\right)^{2}\delta(y).
\end{equation}
The $(i,j)$ components are given by
\begin{center}
\begin{eqnarray}\label{eq:ij}
\left\{-2\frac{\ddot a}{a} -\frac{\ddot b}{b} + \left[\frac{\dot
a}{a} \left(-\frac{\dot a}{a}+2\frac{\dot n}{n} \right)+ \frac{\dot
b}{b} \left(-2\frac{\dot a}{a}+\frac{\dot n}{n}
\right)\right]\right\} \delta_{ij}+\nonumber\\
\left\{\left(\frac{n}{b}\right)^2 \left[2\frac{a^{\prime
\prime}}{a}+\frac{n^{\prime \prime}}{n} + \frac{a^{\prime}}{a}
\left(\frac{a^{\prime}}{a}+2\frac{ n^{\prime}}{n}\right)-
\frac{b^{\prime}}{b} \left(\frac{n^{\prime}}{n}+2\frac{a^{\prime}}
{a}\right) \right]\right\} \delta_{ ij}= \frac{1}{\phi}\left(
\frac{n}{a}\right)^2\left[T^{\phi}_{ij}+T_{ij}+T^{\it
curv}_{ij}\right],
\end{eqnarray}
\end{center}
where
\begin{eqnarray}\label{Teq:ij}
T_{ij}^{\phi}&=&\left\{ \frac{\ddot{\phi}}{\phi} +
\frac{\dot{\phi}}{\phi} \left(2
    \frac{\dot{a}}{a} + \frac{\dot{b}}{b}- \frac{\dot{n}}{n} +
    \frac{\omega}{2} \frac{\dot{\phi}}{\phi} \right) -
  \left(\frac{n}{b}\right)^2 \left[ \frac{\phi^{\prime\prime}}{\phi} +
    \frac{\phi^{\prime}}{\phi} \left(2
    \frac{a^{\prime}}{a}  \frac{b^{\prime}}{b}+ \frac{n^{\prime}}{n} +
    \frac{\omega}{2} \frac{\phi^{\prime}}{\phi} \right)\right]
  \right\} \delta_{ij}
\end{eqnarray}
and
\begin{equation}\label{Tcurveq:ij}
T^{\it
curv}_{ij}=-\frac{1}{\mu^{2}b}\left[\frac{a^{2}}{n^{2}}\delta_{ij}\left(-\frac{\dot{a}^{2}}{a^{2}}+
\frac{\dot{a}}{a}\frac{\dot{n}}{n}-2\frac{\ddot{a}}{a}\right)\right]\delta(y).
\end{equation}
The $(0,5)$ component takes the form
\begin{eqnarray}
\label{eq:05} 3 \left(\frac{\dot{a}}{a}\frac{n^{\prime}}{n} +
\frac{\dot{b}}{b}
  \frac{a^{\prime}}{a} -
  \frac{\dot{a}^{\prime}}{a}\right) =\frac{1}{\phi}T^{\phi}_{05}\,,
\end{eqnarray}
where
\begin{equation}\label{Teq:ijij}
T_{05}^{\phi}=\dot{\phi}^{\prime} - \dot{\phi}\left(
  \frac{n^{\prime}}{n} - \omega\frac{\phi^{\prime}}{\phi}
\right) - \frac{\dot{b}}{b} \phi^{\prime}.
\end{equation}
Finally, for the $(5,5)$ component one has
\begin{eqnarray}
\label{eq:55} 3 \left[-\left(\frac{\ddot a}{a} + \frac{\dot
a}{a}\left(\frac{\dot a}{a} - \frac{\dot n}{n}\right)\right) +
\left(\frac{n}{b} \right)^2 \left(
\frac{a^{\prime}}{a}\left(\frac{a^{\prime}}{a} +
\frac{n^{\prime}}{n}\right)\right) \right]=\frac{1}{\phi}\left(
\frac{n}{b}\right)^2\left[T^{\phi}_{55}+T_{55}\right],
\end{eqnarray}
where
\begin{equation}\label{Teq:55}
T_{55}^{\phi}=\ddot{\phi} +\dot{\phi} \left( 3 \frac{\dot{a}}{a} -
\frac{\dot{n}}{n} + \frac{\omega}{2}
 \frac{\dot{\phi}}{\phi} \right) - \left( \frac{n}{b}\right)^2
\phi^{\prime} \left( 3\frac{a^{\prime}}{a}  + \frac{n^{\prime}}{n} -
\frac{\omega}{2} \frac{\phi^{\prime}}{\phi} \right).
\end{equation}
The equation of motion for the Brans-Dicke field reads
\begin{eqnarray}
\label{eq:bdfield} \ddot{\phi} + \dot{\phi} \left( 3
\frac{\dot{a}}{a} +
  \frac{\dot{b}}{b} - \frac{\dot{n}}{n}\right) - \left( \frac{n}{b}
\right)^2 \left[ \phi^{\prime\prime} + \phi^{\prime} \left( 3
    \frac{a^{\prime}}{a} - \frac{b^{\prime}}{b} + \frac{n^{\prime}}{n}
  \right)\right] & = & - n^2 \frac{\left(T+T^{\it curv}\right)}{3
  \omega+4},
\end{eqnarray}
where a dot represents the time derivative with respect to $\tau$
and the prime corresponds to derivatives with respect to $y$. Note
that in the above derivation we have assumed $\phi=\phi(\tau,y)$.
We make the assumption that the metric and the BD field are
continuous across the brane localized at $y=0$. However, their
derivatives can be discontinuous at the brane position in the $y$
direction. This suggests the second derivatives of the scale
factor and the BD field will have a Dirac delta function
associated with the position of the brane. Since the matter is
localized on the brane it will introduce a delta function in the
Einstein equations which will be matched by the distributional
part of the second derivatives of the scale factor and the BD
field.

Using the Einstein equations it is possible to find out the jump
conditions for $a$ and $n$ by matching the Dirac delta functions
appearing on the left-hand side of the Einstein equations to the
ones coming from the energy-momentum tensor equation (\ref{em0}).
For the BD field one has to use equation (\ref{eq:bdfield}) to
evaluate the jump conditions. We therefore find
\begin{eqnarray}
\label{eq:jumpa}
\frac{[a^{\prime}]_{_{0}}}{a_{_{0}} b_{_{0}}} & = &
-\frac{1}{(3\omega+4)\phi_{_{0}}}\Big[p+ p^{\it curv}+(\omega +1) (\rho+\rho^{\it curv})\Big] \,, \\
\label{eq:jumpn} \frac{[n^{\prime}]_{_{0}}}{n_{_{0}} b_{_{0}}}&=&\frac{1}{(3\omega+4)\phi_{_{0}}}
\Big[3(\omega +1)(p+p^{\it curv})+
(2\omega +3)(\rho+\rho^{\it curv})\Big]\,,\\
\label{eq:jumpbd} \frac{[\phi^{\prime}]_{_{0}}}{\phi_{_{0}}
b_{_{0}}} & = &
\frac{2}{(3\omega+4)\phi_{_{0}}}\Big[\gamma\rho+\gamma_{\rm
c}\rho^{\it curv} \Big]\, ,
\end{eqnarray}
where
\begin{eqnarray}\label{mcurv}
\rho^{\it curv}&=&-\frac{3}{\mu^{2}n^{2}_{_{0}}}\left(\frac{\dot{a}_{_{0}}}{a_{_{0}}}\right)^{2},\\
\label{2} p^{\it curv}&=&\frac{1}{\mu^{2}n^{2}_{_{0}}}\left(\frac{\dot{a}_{_{0}}^{2}}{a_{_{0}}^{2}}-2
\frac{\dot{a}_{_{0}}}{a_{_{0}}}\frac{\dot{n}_{_{0}}}{n_{_{0}}}+2\frac{\ddot{a}_{_{0}}}{a_{_{0}}}\right),
\end{eqnarray}
\begin{eqnarray}\label{gama}
\gamma=\frac{1}{2}(3w_{m}-1)~~~~~~~~~\mbox{and}~~~~~~~~~
\gamma_{c}=\frac{1}{2}(3w_{c}-1),
\end{eqnarray}
with $w_{c}=\frac{p^{\it curv}}{\rho^{\it curv}}$ and the subscript
$0$ stands for the brane at $y=0$. The first two conditions,
equations (\ref{eq:jumpa}) and (\ref{eq:jumpn}), are equivalent to
Israel's junction conditions in general relativity~\cite{israel}
(see \cite{bin} for a discussion of its application in the context
of brane world). It is important to note that the above jump
conditions at $y=0$ depend on the energy density and the pressure
component of the brane world and induced curvature on the brane.
Interestingly, for the radiation dominated phase on the brane,
$\rho=3p$, the jump condition for $\phi$ does not vanish and is
proportional to the energy density and  pressure component of the
induced curvature on the brane.

Taking the jump of the ($0,5$) component of the Einstein equation
and substituting equations (\ref{eq:jumpa} and \ref{eq:jumpn}) we
get the continuity equation for the matter on the brane
\begin{eqnarray}
  \label{eq:cons}
  \dot \rho +3 \left(\rho + p\right) \frac{\dot a}{a}
= 0 \,.
\end{eqnarray}
Equation (\ref{eq:cons}) shows that the energy content of the brane
is still conserved in this scenario which seems to be at odds with
recent results obtained independently by several authors who
conclude that the presence of a dilaton field in the bulk will lead
to a non-trivial coupling with the matter on the brane which, from
the point of view of an observer living on the brane, would be seen
as matter leaking from the brane~\cite{bin1,wands,mennim,mart,ata}.
Our situation is different because the coupling between the BD field
and ordinary matter on the brane was chosen so as to satisfy the
equivalence principle, as was mentioned above. Had we chosen the
coupling in~(\ref{eq:bdfield}) to have the usual $4$-dimensional
value $(2\omega+3)^{-1}$, we would have ended up with a situation
where the conservation equation~(\ref{eq:cons}) would not hold and
energy could leak from the brane.

Taking the mean value of the $(5,5)$ component of Einstein's
equations we can now obtain a Friedmann type equation on the brane
by following a very similar procedure to the one introduced
in~\cite{bin}. Using the fact that due to the orbifold symmetry $y
\leftrightarrow -y$,  the mean value of any of the quantities $a$,
$n$ or $\phi$ should be zero, we can discard all the terms
involving mean values in the average of the $(5,5)$ component of
the Einstein equations. The equation so obtained will still
involve a term containing $\ddot{\phi}$, but we can use the mean
value of the BD field equation~(\ref{eq:bdfield}) to write this in
terms of $a$, $b$ and $n$, and their derivatives. After a somewhat
lengthy calculation we obtain the first Friedmann type equation on
the brane
\begin{eqnarray}
\label{eq:friedtype} \frac{\ddot
a_{_{0}}}{a_{_{0}}}+\left(\frac{\dot a_{_{0}}}{a_{_{0}}}\right)^2+
\frac{\omega}{6}\left(\frac{\dot
\phi_{_{0}}}{\phi_{_{0}}}\right)^2+
\frac{2\gamma\omega}{\mu^{2}(3\omega
+4)}\frac{\dot{a_{_{0}}}^{2}}{a_{_{0}}^{2}\phi_{_{0}}^2}-
\frac{1}{2\mu^{2}(3\omega
+4)}\frac{\dot{a_{_{0}}}^{2}}{a_{_{0}}^{2}\phi_{_{0}}^2}\left[p+\rho+\frac{\omega
}{2}(3p+\rho)\right]\\\nonumber\\\nonumber+
\frac{1}{2}\frac{\dot{a_{_{0}}}^{4}}{a_{_{0}}^{4}\phi_{_{0}}^2}\left[\frac{6\omega}{(3\omega
+4)}+1\right]+ \frac{2\gamma\omega}{\mu^{2}(3\omega
+4)^2}\frac{\ddot a_{_{0}}}{a_{_{0}}\phi_{_{0}}^2}+
\frac{1}{2\mu^{2}(3\omega +4)^2}\frac{\ddot
a_{_{0}}}{a_{_{0}}\phi_{_{0}}^2}(p+\rho+\omega\rho)+\\\nonumber\\\nonumber
\frac{1}{2\mu^{4}(3\omega +4)}\frac{\dot a_{_{0}}^2\ddot
a_{_{0}}}{a_{_{0}}^3\phi_{_{0}}^2}\left[\frac{12\omega}{(3\omega
+4)}-2-3\omega\right]+ \frac{1}{2\mu^{4}(3\omega +4)}\frac{\ddot
a_{_{0}}^2}{a_{_{0}}^2\phi_{_{0}}^2}\left[\frac{6\omega}{(3\omega
+4)}+1\right] =-\frac{\omega}{4 (3\omega +4)^{2}\phi_{_{0}}^2}\\
\nonumber\\\nonumber
\times\left[3p^2+2p\rho+\frac{\rho^2}{3}\right]-\frac{1}{4(3\omega+4)\phi_{_{0}}^2}
\left[\frac{p^2}{2}+p\rho(1+\omega)+\frac{\rho^2}{6}(3+2\omega)\right].\,
\end{eqnarray}
Upon using the components $(5,5)$ and $(0,5)$ we obtain the other
Friedmann type equation
\begin{eqnarray}\label{eq:friedtype1}
\left(\frac{\dot a_{_{0}}}{a_{_{0}}}\right)^2=\frac{1}{(3\omega
+4)^{2}\phi_{_{0}}^{2}}\Big[p^{\it tot}+(\omega +1)\rho^{\it tot}\Big]^{2}-\frac{2}{3a_{_{0}}^{4}}\int \frac{d\tau}{\phi_{_{0}}}\dot{a}_{_{0}}a_{_{0}}^{3}
T^{\phi5}\,_{5}\Big|_{_{y=0}}\\\nonumber\\
\nonumber
-\frac{2}{3a_{_{0}}^{4}}\frac{1}{(3\omega
+4)}\int
\frac{d\tau}{\phi_{_{0}}^{2}}\Big[p^{\it tot}+(\omega +1)\rho^{\it tot}\Big]a_{_{0}}^{4}T^{\phi5}\,_0\Big|_{_{y=0}}.
\end{eqnarray}

While deriving the above equations we have assumed that from the
point of view of the brane observer, the extra dimension is static,
that is $b=b_0$. We have also fixed the time in such a way that
$n_0=1$, corresponding to the usual choice of time in conventional
cosmology. It is interesting to note that by taking the limit
$\omega \rightarrow \infty$ on the right-hand side of equation
(\ref{eq:friedtype}), we obtain exactly the same expression as in
the DGP model \cite{dick2}. Note that in this model $\omega$ is a
coupling constant, but after inducing equations on the brane it can
be interpreted as the induced BD parameter on the brane.

Taking the mean value of the BD field equation we obtain an
equation of motion for $\phi$ on the brane
\begin{eqnarray}
\label{eq:bdeq}
\frac{\ddot\phi_{_{0}}}{\phi_{_{0}}}+3\frac{\dot a_{_{0}}}{a_{_{0}}}\frac{\dot{\phi_{_{0}}}}{\phi_{_{0}}}+\frac{3\omega}{\mu^{2}(3\omega+4)^2
\phi_{_{0}}^{2}}\frac{\ddot a_{_{0}}}{a_{_{0}}}\left[2\gamma+\frac{6}{\mu^{2}}\frac{\dot a_{_{0}}^{2}}{a_{_{0}}^{2}}+
\frac{3}{\mu^{2}}\frac{\ddot{a_{_{0}}}}{a_{_{0}}}\right]\\\nonumber\\\nonumber+\frac{3\omega}{\mu^{2}(3\omega +4)^2
\phi_{_{0}}^{2}}\frac{\dot a_{_{0}}^{2}}{a_{_{0}}^{2}}\left[2\gamma+\frac{3}{\mu^{2}}\frac{\dot{a_{_{0}}}^{2}}{a_{_{0}}^{2}}\right]=
\frac{\omega \gamma^2}{(3\omega +4)^2
\phi_{_{0}}^{2}}\,\,.
\end{eqnarray}
Note that in order to obtain equation (\ref{eq:bdeq}) we also have
to assume that the non-distributional part of $\phi^{\prime\prime}$
vanishes, otherwise, a term involving
$\widehat{\phi^{\prime\prime}}$ will appear in the BD field
equation. As we shall see in the next section it is possible to
obtain cosmologically interesting solutions which verify this
condition. Using  equations (\ref{eq:friedtype}),
(\ref{eq:friedtype1}) and~(\ref{eq:bdeq}) we can determine the
cosmology of the brane at $y=0$.


\section{Cosmological solutions}
It is well know that the original DGP brane cosmology model leads to
self accelerating solutions \cite{deffayet}. In this section, we
start by consider the self accelerating solutions in our model in
the vacuum sector {\it i.e} $\rho=p=0$. Let us take the brane
metric, $q_{\mu\nu}$, as spatially flat in the $4D$ de Sitter
geometry and write
\begin{equation}\label{bm}
q_{\mu\nu}dx^\mu dx^\nu=-d\tau^2+e^{2H\tau} d\mathbf{x}^2.
\end{equation}
Thus, in order to obtain the self accelerating solutions, we make
the {\it ansatz}
\begin{equation}\label{an}
\phi_{_{0}}(\tau)=\alpha a_{_{0}}(\tau),
\end{equation}
where $\alpha$ is an arbitrary positive constant. Using equations
(\ref{eq:friedtype1}), (\ref{bm}) and (\ref{an}) one finds
\begin{equation}\label{self}
H=\epsilon\mu^{2}(3\omega+4)\phi_{_{0}}\sqrt{\frac{\omega+20}{6\omega}},
\end{equation}
where $\epsilon=\pm1$. Note that for $\epsilon=+1$, or the
self-accelerating branch, we will have the self accelerating
solutions in our model.

We now obtain  solutions for $a$ and $\phi$ which will allow us to
discuss the cosmological implications in our setup. Let us then
assume that both the total energy density $\rho$ and pressure $p$ on
the brane consist of two parts
\begin{equation}\label{eqq42}
\rho=\lambda+\varrho~~~~~~~~\mbox{and}~~~~~~~~~p=-\lambda+\textsf{p},
\end{equation}
where $\lambda$, $\varrho$ and $\textsf{p}$ are the tension, the
usual cosmological energy density and pressure in the matter frame
respectively. In what follows we concentrate on the case
$\varrho=\textsf{p}=0$, {\it i.e.} the vacuum solution. However, by
retaining a non-zero effective tension on the brane we are actually
taking the brane effects into account. For simplicity we take the
form of the tension, $\lambda$, as follows \cite{berto}
\begin{equation}\label{tension}
\lambda=\lambda_{c}\phi^{2},
\end{equation}
where $\lambda_c$ is a constant. As before, we assume that the
extra space-like dimension is stabilized ($b$ is constant) and
choose time such that $n_0=1$. We now look for a power law
solution for the scale factor. Substituting the {\it ans\"{a}tze}
\begin{eqnarray}
\label{eq:friedeq}
a_{_{0}}(\tau)\propto \tau^n~~~~~~~~~~\mbox
{and}~~~~~~~~~~~~~~\phi_{_{0}}(\tau)=\frac{\tau^m}{\mu^2},
\end{eqnarray}
into equations (\ref{eq:friedtype}) and (\ref{eq:bdeq}) one
obtains
\begin{equation}\label{eq45}
m=-1,
\end{equation}
and from equation (\ref{eq:bdeq}), $n$ is obtained as the
solution of the following algebraic equation in terms of
$\lambda_{c}$, $\mu^2$ and $\omega$
\begin{eqnarray}\label{eqq46}
\frac{36 \omega  }{(3\omega +4)^2}n^4-\frac{36\omega  }{ (3 \omega +4)^2}n^3+ n^2\left[
\frac{9\omega}{(3\omega +4)^2}-\frac{24 \lambda_{c}\omega }{\mu ^2(3
\omega+4)^2}\right]\\\nonumber\\
\nonumber-n\left[3+\frac{12\lambda_{c}\omega}{\mu^2(3\omega+4)^2}\right]+\frac{4\lambda_{c}^2\omega}{\mu^4(3\omega+4)^2}+2=0.
\end{eqnarray}
This algebraic equation has four explicit real solutions for $n$ in
terms of $\lambda_{c}$, $\mu^2$ and $\omega$.

Equation (\ref{eq:friedeq}) shows that there is a possibility of
having an accelerated expanding universe for some choices of
$\omega$, $\mu^2$ and $\lambda_{c}$. The deceleration parameter on
the brane as a function of $\omega$, $\mu^2$ and $\lambda_{c}$ is
given by
\begin{equation}\label{eq49}
q(\omega,\lambda_{c},\mu^2)=-\frac{a_{_{0}}\ddot{a}_{_{0}}}{\dot{a}_{_{0}}^{2}}=-\frac{n-1}{n},
\end{equation}
where $n$ is one of the four roots of equation (\ref{eqq46}).  A
glance at equation (\ref{eq49}) reveals that using the condition for
acceleration, that is $q(\omega,\lambda_{c},\mu^2)<0$, leads to
$n>1$ from which, using definition $w_{\rm
eff}=-1-\frac{2\dot{H}_{_{0}}}{3H_{_{0}}^2}$ for the effective
quintessence, we find $w_{\rm eff}<-1/3$. Figure 1  shows the
behavior of the deceleration parameter, $q$, as a function of
$\omega$ for $\frac{\lambda_{c}}{\mu^2}\sim1$ and $n$, taken as a
root of equation (\ref{eqq46}). Therefore, each graph in this figure
corresponds to one of the roots of the fourth order algebraic
equation (\ref{eqq46}). As can be seen, the graphs in the first row
in Figure 1 show that for positive and negative values of $\omega$,
we have an accelerating universe. The graph on the top left hand
corner is particularly interesting since it shows that for
$\omega\longrightarrow\pm\infty$ we have $q\rightarrow-1$, that is
the universe finally approaches the eternal de Sitter phase. It has
been claimed that the value of $|\omega|$ in $4D$ should be large
$(> 40000)$ if Brans-Dicke theory is to be consistent with the
astronomical observations \cite{bertotti}. Therefore, the graph
mentioned above seems to be in good agreement with the observational
data. The graphs in the second row  in figure 1 predict a
decelerating universe which is not consistent with the present
observational data. However, the model can be used to interpolate
back the deceleration parameter to an earlier epoch to yield a
decelerating universe which is required in order to explain
processes like nucleosynthesis.

\begin{figure}
\begin{center}
\epsfig{figure=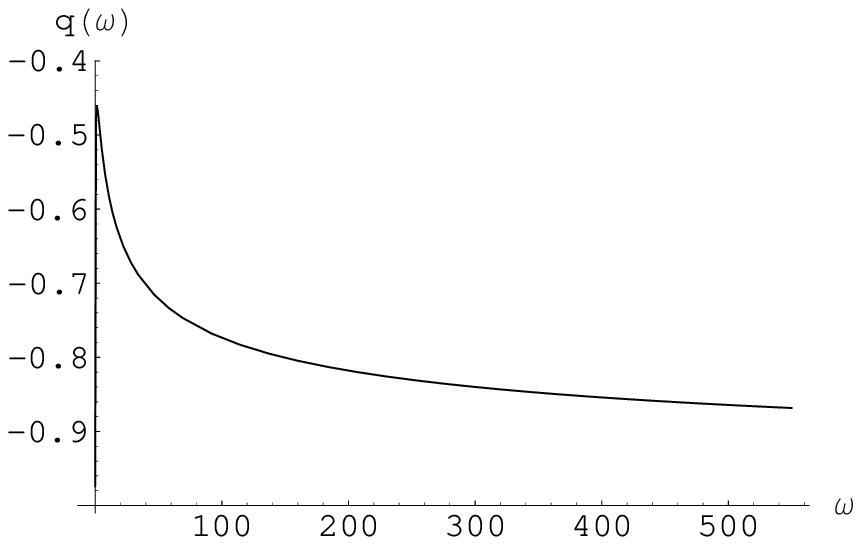,width=8cm}\hspace{5mm}
\epsfig{figure=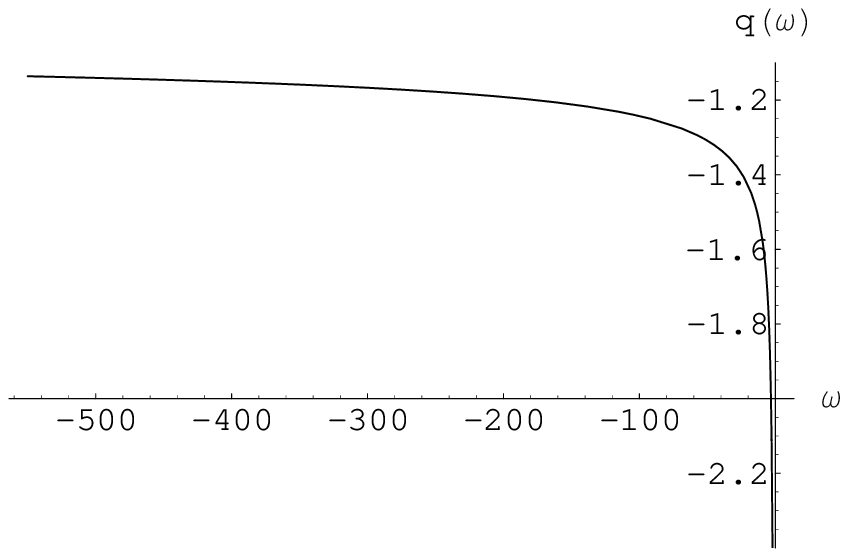,width=8cm}\vspace{5mm}
\epsfig{figure=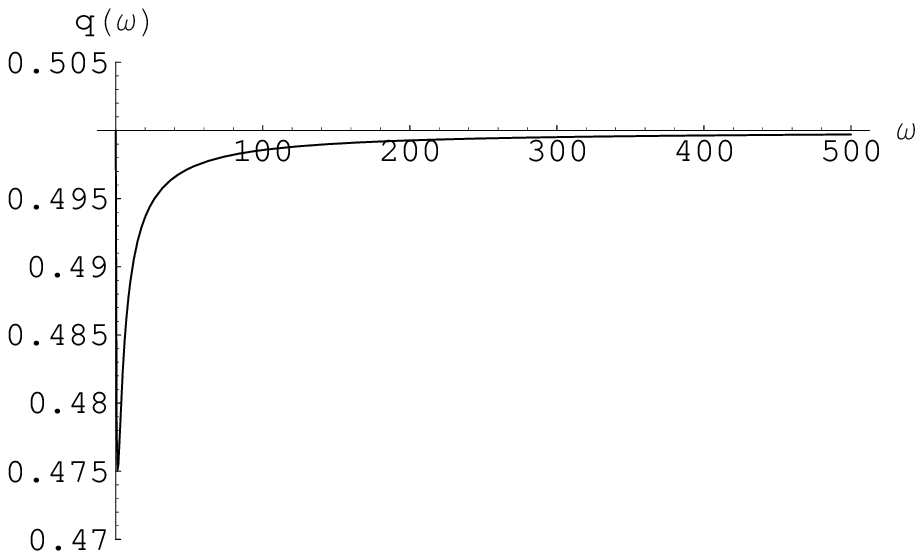,width=8cm}\hspace{5mm}
\epsfig{figure=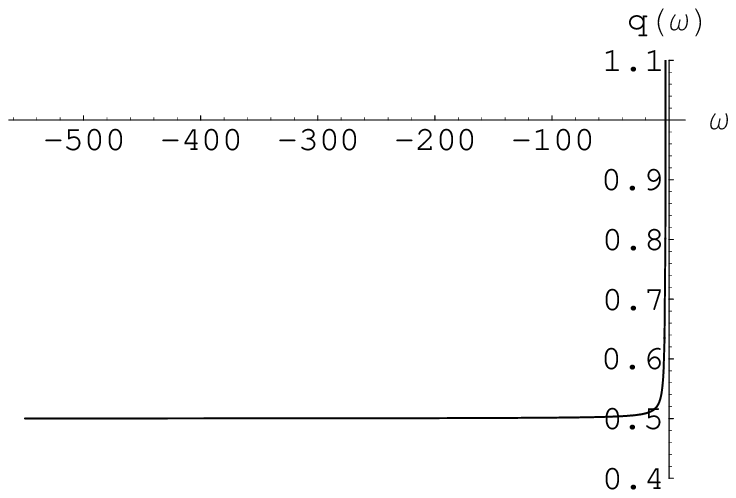,width=8cm}
\end{center}
\caption{\footnotesize  The behavior of $q(\omega)$ as a function
of $\omega$ for $\frac{\lambda_c}{\mu^2}\sim1$. Each figure
represents the deceleration parameter for one of the four roots of
equation (\ref{eqq46}).}
\end{figure}
\section{Conclusions}
We have derived the effective four dimensional field equations in a
DGP brane-world setting where a BD field is assumed to be present in
the bulk. We considered the weak field approximation in our model
which resulted in the modification of the mass appearing in the
gravitational potential by the addition of the mass of the effective
intrinsic curvature that appears on the brane. In our scenario, the
conservation equation for the matter fields confined to the brane
still holds in spite of the existence of a BD field in the bulk.
This is due to the fact that our calculations were done in the
Jordan frame. Finally we obtained the modified Friedmann equation on
the brane and showed that an accelerated expanding universe results
for certain choices of the parameters which is in good agreement
with the present bounds on the value of $\omega$ in Brans-Dicke
theory.


\end{document}